# Effect of mixed alkali-element substitution on structural and magnetic properties of praseodymium manganites $Pr_{0.9}(Na_{1-x}K_x)_{0.1}MnO_3$

S. Zouari[a,b], L. Ranno[b], A. Cheikh-Rouhou[a] and P. Strobel[b†]

[a]*Laboratoire de Physique des Matériaux, Faculté des Sciences de Sfax, B. P. 763, 3038 Sfax, Tunisia.*

[b]*Institut Néel, CNRS, B. P. 166, 38042 Grenoble, Cedex 9 France.*

**Abstract**

The effect of cationic size mismatch at the A site at constant manganese valence on the structural and magnetic properties of a perovskite-type rare-earth manganate was investigated in the $Pr_{0.9}(Na_{1-x}K_x)_{0.1}MnO_3$ solid solution system ($0 \leq x < 1$). All members of this solid solution series are orthorhombic at room temperature, space group Pbnm. Structural refinements using the Rietveld method show that the cell volume increases and the static Jahn-Teller distortion decreases with increasing potassium content x. Magnetic properties are characterized by strong positive $\theta_p$ values, and are ascribed to a canted ferromagnetic arrangement. The high Curie constant values in the paramagnetic regime can be explained by a magnetic cluster model of 2-3 Mn ions.

† Corresponding author

Postal address: Institut Néel, CNRS, BP 166, case F, 38042 Grenoble Cedex 9, France

Phone: 33+ (0)476 887 940, Fax 33+ (0)476 881 038

e-mail: pierre.strobel@grenoble.cnrs.fr

**Introduction**

Perovskite-type manganese oxides with general formula $Ln_{1-x}A_xMnO_3$, have been extensively studied for their remarkable magnetoresistive properties [1-4]. Such properties are of a considerable interest for a new generation of magnetic sensors and recording devices. The magnetic behaviour in these systems is very sensitive to the manganese valence, and it can be actually tuned by adjusting the fraction x of substituent A, usually calcium or strontium. Much fewer studies are available on systems containing alkali elements as A substituents to modify the charge distribution in these materials; we are aware of a few studies with Ln = La, Pr or Nd and A = Na or K [5-10]. Yet sodium and potassium also have appropriate sizes to occupy the A site of the perovskite structure.

Another important parameter influencing the magnetic properties of perovskite-type manganese oxides has been shown to be the size of the A cation [11, 12]. In order to study this parameter independently, many groups investigated $Ln_{1-x}A_xMnO_3$ systems where A itself is a combination of elements, in general $(Sr_xCa_{1-x})$. Again, the equivalent systems with alkali metal components have been the subject of very few studies, in spite of the fact that they permit a much wider variation in size than the Ca-Sr couple: the ionic radii for 12-fold coordination vary between 1.34 and 1.44 Å for the Ca-Sr couple, compared to 1.39 and 1.64 Å for the Na-K couple [13]. In the alkaline-earth substituted systems, larger ionic radii can be supplied using barium, but to our knowledge no study of combined $Ln_{1-y}(Sr_{1-x}Ba_x)_yMnO_3$ is available. In the $La_{0.85}(Na_{1-x}K_x)_{0.15}MnO_3$ system, the Curie temperature at constant doping level was found to be independent of the Na/K ratio [14]. In the equivalent praseodymium system, on the other hand, the magnetic properties have been found to be composition-dependent; they have been analyzed in terms of magnetic clusters above $T_c$, the size of which depends on the Na/K ratio [15]. These studies show the interest of $Ln_{1-x}A_xMnO_3$ compounds containing both sodium and potassium. This paper thus addresses the structural and magnetic properties in the system $Pr_{0.9}(Na_{1-x}K_x)_{0.1}MnO_3$.

**Experimental**

Powder samples of $Pr_{0.9}(Na_{1-x}K_x)_{0.1}MnO_3$ with various values of x have been prepared using the solid state reaction of appropriate amounts of reagent grade praseodymium oxide ($Pr_6O_{11}$), manganese dioxide ($MnO_2$), sodium and potassium carbonates ($K_2CO_3$). The reagents were mixed and ground with mortar and pestle, and fired repeatedly for 36 hours at 1000°C, then at 1150°C in air with intermediate regrinding. Samples were furnace-cooled to at least 300°C before removal, in order to prevent the formation of oxygen vacancies.

Phase purity, homogeneity and unit cell dimensions were determined by powder X-ray diffraction at room temperature in transmission geometry, using a Bruker D8 diffractometer equipped with Cu-K*a* radiation and an incident-beam monochromator. Diffractograms were recorded in the range 21–81° using a step size of 0.02 ° and counting rate 100 s per step. Structures were refined using room-temperature X-ray data by the Rietveld method, using the Fullprof program [16]. Additional chemical analyses were carried out by EDS in a JEOL 840 scanning electron microscope (SEM).

Magnetization measurements versus temperature with magnetic field up to 8 T were recorded by an extraction magnetometer in the temperature range 2–300 K.

**Results and discussion**

*1. Crystal chemistry*

All samples gave X-Ray diffraction patterns indexable in an orthorhombically-distorted perovskite-type structure. All samples contained a weak impurity identified as apatite-type praseodymium silicate $Pr_{9.33}Si_6O_{26}$ (in apatite-structure notation $Pr_{4.66}(SiO_4)_3O$ [17] ). The formation of traces of a silicate impurity may be due to slight contamination in the furnace during heat treatments. This phase was included in Rietveld refinements (see Fig.1). Its fraction was found ≤ 1.5 % in all cases. These results were confirmed by SEM-EDS analysis. The samples consisted of isotropic grains with diameter in the range 2-5 μm. Global analyses on different samples yielded a fairly constant overall silicon contamination (ca. 1.2 at.% Si), whereas analyses on individual grains yielded either no silicon or much higher silicon contents on a few specific grains. This is consistent with the presence of isolated grains of a second phase. The influence of the apatite phase on the main perovskite phase stoichiometry was checked by freeing the praseodymium or manganese occupation in Rietveld refinements: the shifts from full occupation were found to be negligible (within standard deviations) in both cases. Given the small concentration of the impurity, the absence of manganese in it, and the negligible effect on the perovskite stoichiometry, we assume that the apatite phase does not have any significant effect on the subsequent measurements of magnetic properties.

The main results of structural refinements are summarized in Table 1. A typical X-ray pattern, showing calculated and observed profiles, is given in Figure 1. The Pbnm space group was used, with cell parameters in the order $c/\sqrt{2}<a<b$, which is characteristic of a static O'-type Jahn-Teller deformation in $Mn^{3+}$-containing perovskites [18]. This system differs significantly from $La_{0.85}(Na_{1-x}K_x)_{0.15}MnO_3$ and $Pr_{0.85}(Na_{1-x}K_x)_{0.15}MnO_3$ [15], where all (La) or the major part (Pr) of compositions have $a \approx b \approx c/\sqrt{2}$, and can therefore be described in a rhombohedral perovskite structure with $R\bar{3}c$ space group. In the $Pr_{0.9}A_{0.1}MnO_3$ system, the cell parameter b remains significantly larger than a and $c/\sqrt{2}$ over the whole range of compositions (see Figure 2). The net result is a monotonous cell volume increase with increasing potassium content. As shown in Fig.2, the main expansion occurs along the y-axis (*b* cell parameter), whereas *a* and *c* exhibit very small (and parallel) increase with potassium content. The overall variation is consistent with the increase in average ionic radius when replacing $Na^+$ by $K^+$.

Mn-O distances are given in Table 2. The Jahn-Teller effect is rather small, as shown previously for potassium-only substitution : unsubstituted $PrMnO_3$ exhibits two typical long Mn-O distances of 2.16-2.19 Å, which are suppressed for 0.1 K substitution [6]. This trend is confirmed here; the longest distance observed is 2.025 Å (for x = 0.6). The angles in the $MnO_6$ octahedra also show a limited distortion. Whereas most adjacent O-Mn-O angles in $Pr_{0.9}Na_{0.1}MnO_3$ are 85.5 and 94.5°, this distortion decreases with the introduction of potassium : for $0.2 \leq x \leq 0.6$, all adjacent angles lie within 1.7° from right angles. The distortion increases again at higher potassium content; this trend is especially visible in the evolution of the O1-Mn-O1(a) angle, which goes from 85/95° to 90° for x = 0.4 then to $\approx$ 86/94° for x = 0.8 (see Table 2).

*2. Magnetic properties*

The evolution of magnetization versus magnetic applied field for several temperatures is given in Fig. 3 for x = 0.2 . The magnetization is saturated in 1 T field but a non-negligible high field susceptibility $\approx 0.1$ $\mu_B/T$ is observed. The M(H) curve at 5 K for $Pr_{0.9}(Na_{0.8}K_{0.2})_{0.1}MnO_3$ is compared to that in the previously studied composition $Pr_{0.85}(Na_{0.8}K_{0.2})_{0.15}MnO_3$ at 5 K in Fig. 4, showing that M is smaller for the higher substitution level (0.15), in agreement with the decrease in spin number of $Mn^{4+}$ compared to $Mn^{3+}$.

Fig. 5 shows the evolution of spontaneous magnetization at low temperature and of the inverse susceptibility ($\chi^{-1}$) extracted from the magnetic isotherms as a function of temperature. The measured spontaneous magnetization at T = 5 K is about 3.5 $\mu_B$/f.u., compared to 3.8 $\mu_B$/f.u. for the spin-only magnetic moment as expected in a manganese oxide where the Mn orbital moment is quenched. Together with the high field susceptibility, this could indicate a slightly canted ferromagnetic structure or an antiparallel contribution for $Pr^{3+}$ ions [10]. However, M(T) data at low temperature do not show any significant anomaly as expected for the occurence of a long-range magnetic of $Pr^{3+}$ ions (Fig.5). In addition, the saturation magnetization is governed only by the $Mn^{3+}/Mn^{4+}$ ratio, which is constant in this series and almost independent of the Na/K ratio. Such behavior has been observed previously in $Pr_{0.85}(Na_{1-x}K_x)_{0.15}MnO_3$, with a $Mn^{4+}$ content of 30% [15].

The inverse susceptibility curves exhibit a quasi Curie-Weiss behaviour at high temperature, and can be satisfactorily fitted to linear variation above 220 for all samples. Curie temperatures $T_C$ and extrapolated Weiss temperatures $\theta_p$ values (see Table 3) are close to those reported previously for the $Pr_{0.85}(Na_{1-x}K_x)_{0.15}MnO_3$ series [15]. The increase in $T_C$ with potassium substitution can be correlated with the decrease in Mn-O octahedron distortion – see especially the evolution of O-Mn-O bonding angles in Table 2. $T_C$ indeed goes through a maximum for intermediate (Na, K) compositions, just as the bonding angles get closer to 90°C for mixed (Na, K) compositions. The Weiss temperatures $\theta_p$ are all widely positive (between 140 and 150 K), indicating that the dominating interactions are ferromagnetic. $Mn^{3+}$ and $Mn^{4+}$, present in the $Pr_{0.9}(Na_{1-x}K_x)_{0.1}MnO_3$ series with average manganese valence equal to 3.20+, have spin-only moments given by g = 2 and S = 2 or S = 3/2. The measured inverse susceptibility curves give Curie constant values that are much higher than that expected ones from paramagnetic susceptibility of free manganese ion, indicating the presence of magnetic clusters. Some possible clusters of $Mn^{3+}$ and $Mn^{4+}$ together are defined in the Curie constant

$$C_{cl} = (\frac{N}{n})(\frac{\mu_B}{3k_B})g^2 S_{cl}(S_{cl}+1),$$

where $S_{cl} = n(0.8 S_{Mn^{3+}} + 0.2 S_{Mn^{4+}})$.

According to this model, and using the experimental C(T) values, it is possible to deduce the size of the manganese cluster *n* as a function of temperature. Fig. 6 shows that the *n* value decreases when T increases. The size of the cluster is about to 2-3 Mn ions. This value is smaller than in the $Pr_{0.85}(Na_{1-x}K_x)_{0.15}MnO_3$ system (5-8 Mn ions) [15]. The difference may be explained by the increase in $Mn^{4+}$ content in the latter series.

**Conclusions**

The solid solution systems $Ln_{1-y}(Na_{1-x}K_x)_yMnO_3$. allow to study the physical properties of perovskite-type manganese oxides with the A-site size as only variable, while keeping the manganese valence constant through a fixed value of y. The mixed alkali substitution allows a much wider average size variation on the a-site than the more widely studied alkaline-earth systems. Using a rather small doping level (y = 0.10), we confirm here that potassium depresses the static Jahn-Teller distortion in the praseodymium system. Consistent with previous studies on the Pr-K-Mn-O system with variable manganese valence [6, 10], this study confirms that alkali element substitution in Pr manganites induces a transition from antiferromagnetism to canted ferromagnetism. The uncommon values of Curie constants are consistent with a cluster model, with a number of Mn atoms in the cluster equal to 2-3. This is a consistent with previous results on the same system with higher doping level (y = 0.15), which correspond to larger clusters [15].

**Acknowledgments**

The authors gratefully acknowledge Région Rhône-Alpes for financial support of S. Zouari (MIRA Program).

**References**

1. S. Jin, T.H.Tiefel, M. McCormack, R.A. Fastnacht, R. Ramesh, L.H. Chien, *Science* **264** (1994) 413-415.

2. J. M. D. Coey, M. Viret, S. von Molnar, *Adv. Phys.* **48** (1999) 167-293.

3. B. Raveau, A. Maignan, C. Martin, M. Hervieu, *Chem. Mater.* **10** (1998) 2641-2652.

4. C.N.R. Rao, R. Mahesh, A.K. Raychaudhuri, R. Mahendiran, *J. Phys. Chem. Solids* **59** (1998) 487-501.

5. M. Itoh, T. Shimura, J.D. Yu, T. Hayashi, Y.F. Inaguma, A. Damay, C. Maignan, B. Martin, B. Raveau, *Phys. Rev. B* **52** (1995) 12522-12530.

6. Z. Jirak, J. Hejtmanek, K. Knizek, R. Sonntag, *J. Solid. State. Chem.* **132** (1997) 98-106.

7. G.H. Rao, J.R. Sun, K. Barner, N. Hamad, *J. Phys. Cond. Matt.* **11** (1999) 1523-1528.

8. C. Shivakumara, M.S. Hegde, T. Srinivasa, N.Y. Vasanthacharya, G.N. Subbanna, N.P. Lalla, *J. Mater. Chem.* **11** (2001) 2572-2579.

9. Z. Jirak, J. Hejtmanek, K. Knizek, M. Marysko, E. Pollert, M. Dlouha, S. Vratislav, M. Hervieu, *J. Magn. Magn. Mater.* **250** (2002) 275-287

10. S. Zouari, A. Cheikh-Rouhou, P. Strobel, M. Pernet, J. Pierre, *J. Alloys. and Compounds*. **333** (2002) 21-27.

11. H.Y. Hwang, S.W. Cheong, P.G. Radaelli, M. Marezio, B. Batlogg, *Phys. Rev. Lett*. **75** (1995) 914-917.

12. F. Damay, C. Martin, B. Raveau, *J. Appl. Phys*. **82** (1997) 6181-6185

13. R. D. Shannon, *Acta. Crystallogr. Sect. A*, **32** (1976) 751-767.

14. Z. El-Fadli, E. Coret, F. Sapina, E. Martinez, A. Beltran, D. Beltran, F. Lioret, *J. Mater. Chem.* **9** (1999) 1793-1799.

15. S. Zouari, L. Ranno, A. Cheikh-Rouhou, M. Pernet, P. Strobel, *Solid. State. Com.* **119** (2001) 517-521.

16. J. Rodriguez-Carvajal, *Physica. B.* **192** (1993) 55-69.

17. D.A. Grisafe, F.A. Hummel, *Amer. Mineralogist*, **55** (1970) 1131-1145.

18. J. B. Goodenough, J. Longo, Landolt-Börnstein *Numerical Data and Functional Relationships in Science and Technology*, group III, Vol. 4, Berlin, Springer, 1970.

**Figure Captions**

Fig. 1. Observed (points) and calculated (continuous line) X-ray powder diffraction patterns for $Pr_{0.9}(Na_{0.8}K_{0.2})_{0.1}MnO_3$. The lower part shows the difference Iobs–Icalc. Vertical bars indicate the positions of reflections of the perovskite phase (upper row) and of the apatite-type phase (lower row).

Fig. 2. Evolution of cell parameters with x along the $Pr_{0.9}(Na_{1-x}K_x)_{0.1}MnO_3$ series (a: triangles up, b squares, $c/\sqrt{2}$ triangles down).

Fig. 3. Evolution of magnetization M versus applied magnetic field at different temperatures for $Pr_{0.9}(Na_{0.8}K_{0.2})_{0.1}MnO_3$.

Fig. 4. Comparison of M(H) curves at 5 K for two samples with same Na/K ratio and different average Mn valences : $Pr_{0.9}(Na_{0.8}K_{0.2})_{0.1}MnO_3$ (circles) and $Pr_{0.85}(Na_{0.8}K_{0.2})_{0.15}MnO_3$ (squares).

Fig. 5. Temperature dependence of the saturated magnetization and the inverse susceptibility for $Pr_{0.9}(Na_{1-x}K_x)_{0.1}MnO_3$.

Fig. 6. Evolution of the number of Mn ions per cluster n versus temperature for various members of the $Pr_{0.9}(Na_{1-x}K_x)_{0.1}MnO_3$ series.

Table 1. Results of Rietveld refinements for various $Pr_{0.9}(Na_{1-x}K_x)_{0.1}MnO_3$ compositions. Space group Pbnm.

| x | | 0 | 0.2 | 0.4 | 0.6 | 0.8 |
|---|---|---|---|---|---|---|
| *Cell parameters (Å)* | | | | | | |
| | a | 5.4590(2) | 5.4587(2) | 5.4594(2) | 5.4623(3) | 5.4645(4) |
| | b | 5.4666(2) | 5.4749(2) | 5.4764(2) | 5.4899(3) | 5.5002(4) |
| | c | 7.7154(3) | 7.7181(3) | 7.7191(3) | 7.7235(3) | 7.7231(5) |
| *Positional and thermal parameters* | | | | | | |
| Pr,Na,K | x | 0.997(2) | 0.995(1) | 0.995(1) | 0.994(1) | 0.995(1) |
| (4c) | y | 0.0271(4) | 0.0298(3) | 0.030(3) | 0.032(3) | 0.032(4) |
| | z | 0.25 | 0.25 | 0.25 | 0.25 | 0.25 |
| | $B(Å^2)$ | 0.65(4) | 0.69(4) | 0.70(3) | 0.52(4) | 0.58(5) |
| Mn (4b:) | x, | 0.5 | 0.5 | 0.5 | 0.5 | 0.5 |
| | y | 0 | 0 | 0 | 0 | 0 |
| | z | 0 | 0 | 0 | 0 | 0 |
| | $B(Å^2)$ | 0.28(5) | 0.20(5) | 0.26(5) | 0.41(5) | 0.41(7) |
| O1 (4c) † | x | 0.089(7) | 0.069(6) | 0.074(5) | 0.079(5) | 0.090(6) |
| | y | 0.482(4) | 0.488(3) | 0.488(3) | 0.485(3) | 0.473(4) |
| | z | 0.25 | 0.25 | 0.25 | 0.25 | 0.25 |
| O2 (8d) † | x | 0.722(6) | 0.717(5) | 0.724(5) | 0.721(4) | 0.723(5) |
| | y | 0.288(5) | 0.285(3) | 0.287(3) | 0.292(3) | 0.269(5) |
| | z | 0.027(4) | 0.039(3) | 0.040(3) | 0.036(3) | 0.038(3) |
| *Statistical parameters* | | | | | | |
| | N-P+C | 2905 | 2828 | 2816 | 2828 | 2856 |
| | $R_{wp}$ | 7.44 | 8.24 | 7.82 | 8.44 | 10.4 |
| | $R_{exp}$ | 6.53 | 3.86 | 3.89 | 7.61 | 5.06 |
| | $\chi^2$ | 3.68 | 4.56 | 4.04 | 3.82 | 9.52 |
| | $R_{Bragg}$ | 2.12 | 2.89 | 3.21 | 4.45 | 5.10 |

† B fixed to 1.0.

Table 2. Main Mn-O distances and angles throughout the $Pr_{0.9}(Na_{1-x}K_x)_{0.1}MnO_3$ series

| *Interatomic distances (Å)* | | | | | |
|---|---|---|---|---|---|
| Mn-$O_1$ (×2) | 1.993(5) | 1.971(3) | 1.973(5) | 1.980(4) | 1.998(4) |
| Mn-$O_2$ (×2) (a) | 1.921(14) | 1.936(7) | 1.930(9) | 1.925(9) | 1.939(15) |
| Mn-$O_2$ (×2) (b) | 1.998(13) | 2.005(7) | 2.015(9) | 2.025(9) | 1.997(15) |
| *Interatomic angles (°)* | | | | | |
| O1-Mn-O1 | 180 | 180 | 180 | 180 | 180 |
| O1-Mn-O2 (a) | 85.0/95.0(8) | 89.8/90.2(6) | 90.0(6) | 88.3/91.7(6) | 86.2/93.8(8) |
| O1-Mn-O2 (b) | 86.6/93.4(9) | 89.3/90.7(6) | 89.1/90.9(7) | 89.6/90.4(7) | 89.9/90.1(9) |
| O2-Mn-O2 | 89.1/90.9(9) | 88.6/91.4(5) | 88.5/91.5(7) | 88.7/91.3(7) | 89.3/90.7(10) |

Table 3. Curie and Weiss constants for various compositions

| sample composition | $T_c/K$ | $\theta_p/K$ |
|---|---|---|
| $Pr_{0.9}(Na_{0.8}K_{0.2})_{0.1}MnO_3$ | 130 | 154 |
| $Pr_{0.9}(Na_{0.6}K_{0.4})_{0.1}MnO_3$ | 135 | 150 |
| $Pr_{0.9}(Na_{0.4}K_{0.6})_{0.1}MnO_3$ | 137 | 145 |
| $Pr_{0.9}(Na_{0.2}K_{0.8})_{0.1}MnO_3$ | 140 | |
| $Pr_{0.9}K_{0.1}MnO_3$ [6] | 125 | |

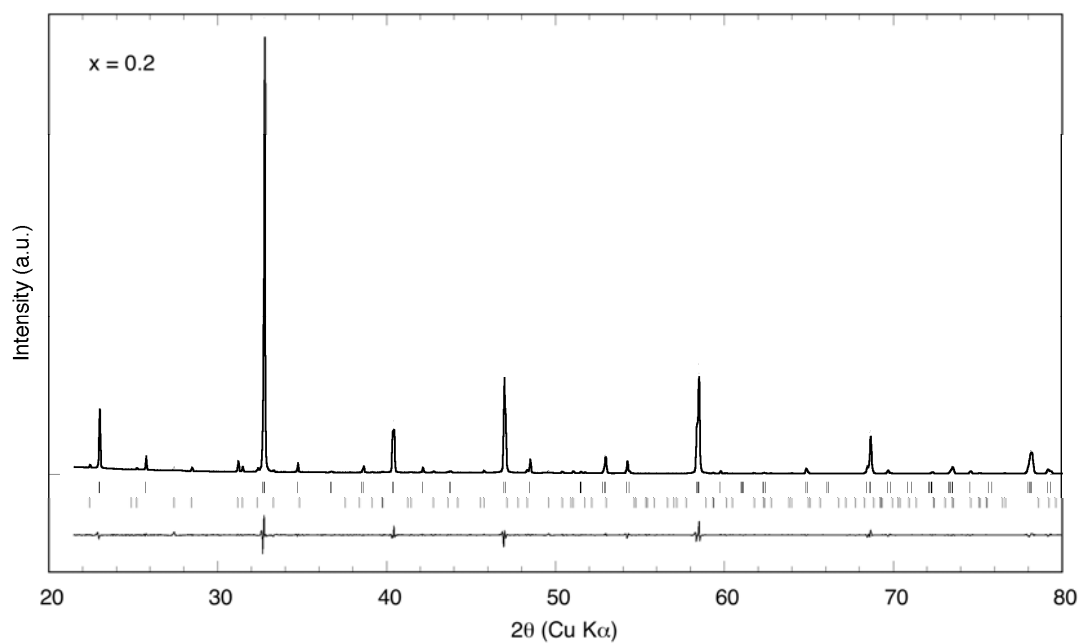

Fig. 1. Observed (points) and calculated (continuous line) X-ray powder diffraction patterns for $Pr_{0.9}(Na_{0.8}K_{0.2})_{0.1}MnO_3$. The lower part shows the difference Iobs–Icalc. Vertical bars indicate the positions of reflections of the perovskite phase (upper row) and of the apatite-type phase (lower row).

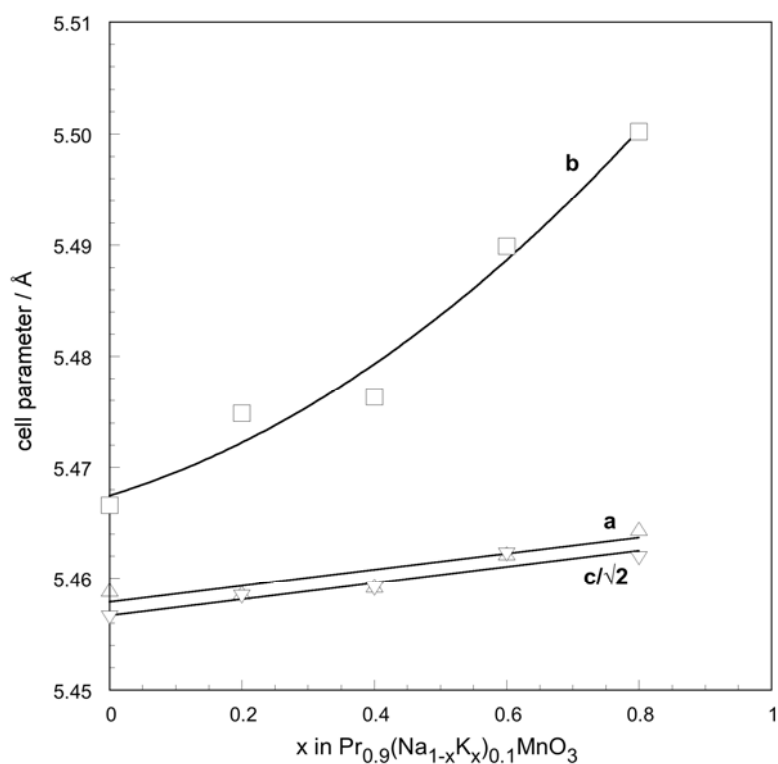

Fig. 2. Evolution of cell parameters with x along the $Pr_{0.9}(Na_{1-x}K_x)_{0.1}MnO_3$ series (a: triangles up, b squares, $c/\sqrt{2}$ triangles down).

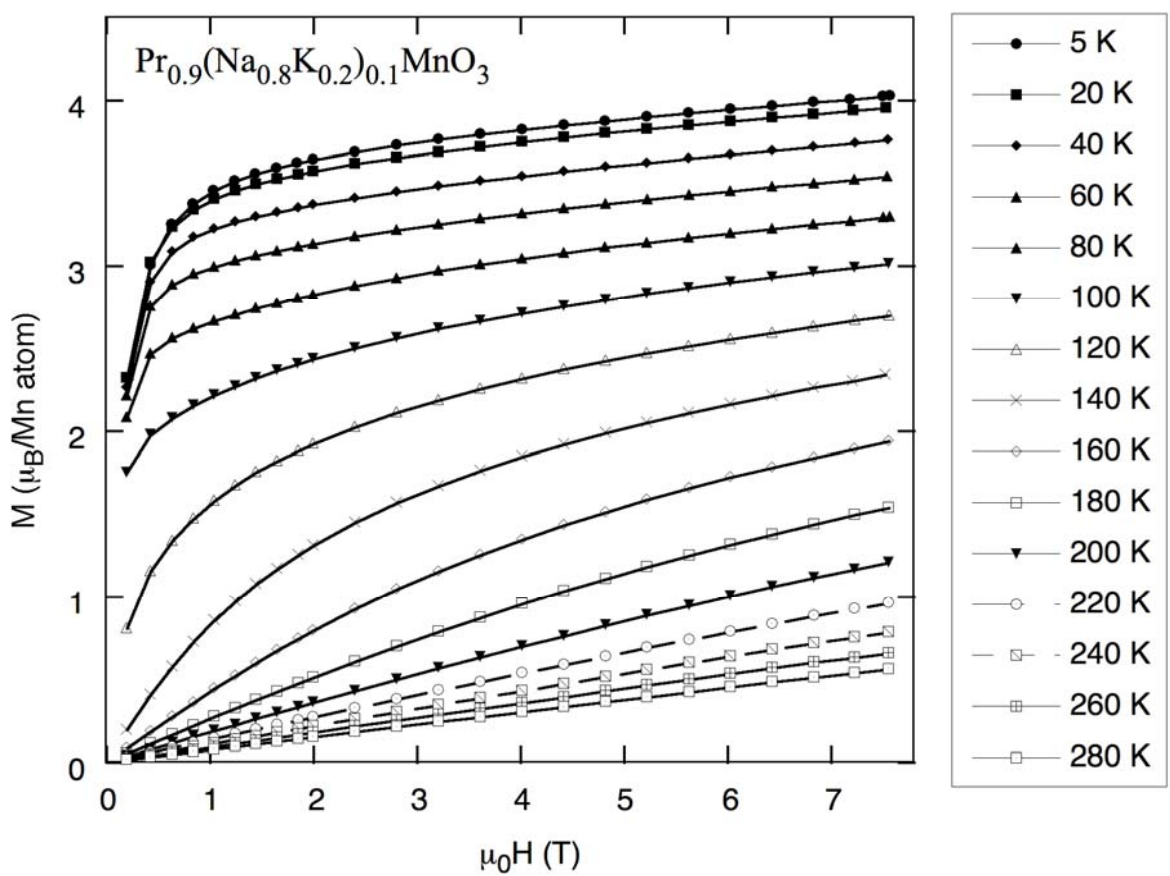

Fig. 3. Evolution of magnetization M versus applied magnetic field at different temperatures for $Pr_{0.9}(Na_{0.8}K_{0.2})_{0.1}MnO_3$.

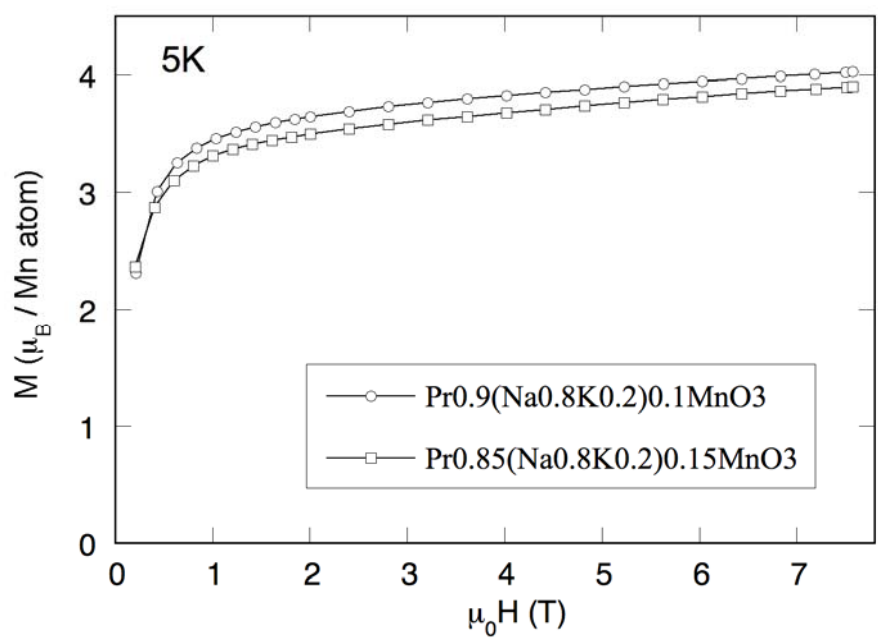

Fig. 4. Comparison of M(H) curves at 5 K for two samples with same Na/K ratio and different average Mn valences : $Pr_{0.9}(Na_{0.8}K_{0.2})_{0.1}MnO_3$ (circles) and $Pr_{0.85}(Na_{0.8}K_{0.2})_{0.15}MnO_3$ (squares).

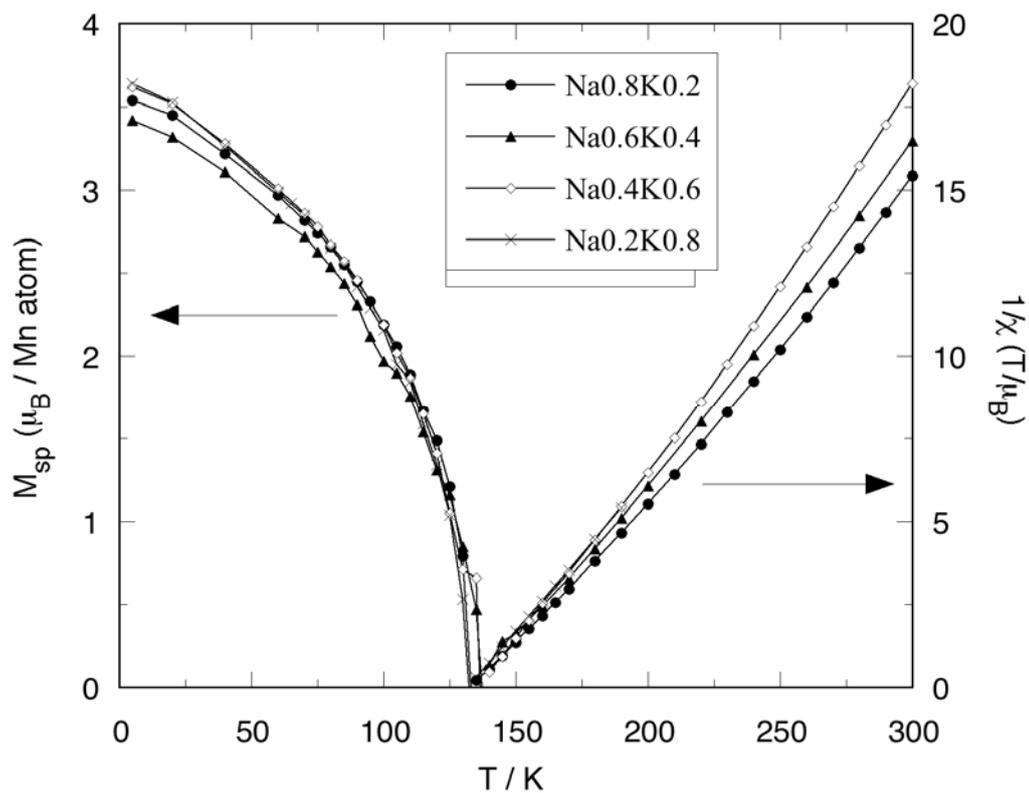

Fig. 5. Temperature dependence of the saturated magnetization and the inverse susceptibility for $Pr_{0.9}(Na_{1-x}K_x)_{0.1}MnO_3$.

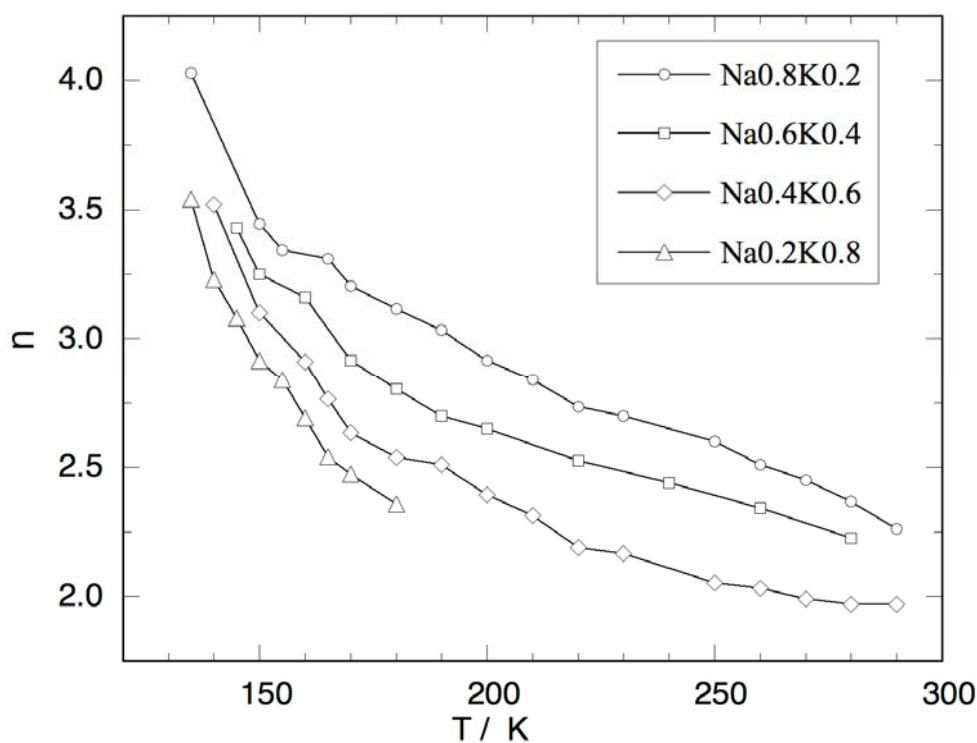

Fig. 6. Evolution of the number of Mn ions per cluster n versus temperature for various members of the $Pr_{0.9}(Na_{1-x}K_x)_{0.1}MnO_3$ series.